\catcode`\@=11
\def\gsim{\ifmmode{\mathrel{\mathpalette\@versim>}}
    \else{$\mathrel{\mathpalette\@versim>}$}\fi}
\def\lsim{\ifmmode{\mathrel{\mathpalette\@versim<}}
    \else{$\mathrel{\mathpalette\@versim<}$}\fi}
\def\@versim#1#2{\lower 2.9truept \vbox{\baselineskip 0pt \lineskip 
    0.5truept \ialign{$\m@th#1\hfil##\hfil$\crcr#2\crcr\sim\crcr}}}
\catcode`\@=12
\def\arccos{{\rm arccos}}
\def\arccosh{{\rm arccosh}}
\def\arctan{{\rm arctan}}
\def\arctanh{{\rm arctanh}}
\def\As{A_1}
\def\Ah{A_0}
\def\Ass{A_{11}}
\def\Ash{A_{10}}
\def\Ahh{A_{00}}
\def\Ahs{A_{01}}
\def\eps{{\cal E}}
\def\epu{\epsilon_1}
\def\epd{\epsilon_2}
\def\fid{f_{\rm k}}
\def\ftil{\tilde f}
\def\Ftil{\tilde F}
\def\Ftilis{\Ftil_{\rm i}}
\def\Ftilan{\Ftil_{\rm a}}
\def\fn{f_{\rm N}}
\def\fis{f_{\rm i}}
\def\fan{f_{\rm a}}
\def\fst{f_1}
\def\fha{f_0}
\def\gau{\gamma_1}
\def\gad{\gamma_2}
\def\Is{I_1}
\def\Ih{I_0}
\def\Iss{I_{11}}
\def\Ish{I_{10}}
\def\Ihh{I_{00}}
\def\Ihs{I_{01}}
\def\ln{\hbox{${\rm ln}\, $}}

\def\parn{\par\noindent}
\def\psih{\Psi_0}
\def\psis{\Psi_1}
\def\psin{\Psi_{\rm N}}
\def\psit{\Psi_{\rm T}}
\def\psiu{\Psi_{\gau}}
\def\psid{\Psi_{\gad}}
\def\psii{\Psi_{\rm k}}

\def\psitilh{\tilde\psih}
\def\psitils{\tilde\psis}
\def\qi{Q_{\rm k}}
\def\mh{M_0}
\def\ms{M_1}
\def\mt{M_{\rm T}}
\def\Mu{M_{\gau}}
\def\Md{M_{\gad}}
\def\nucr{\nu_{\rm cr}}
\def\psitilt{\tilde\psit}
\def\Qtil{\tilde Q}
\def\ra{r_{\rm a}}
\def\rai{r_{\rm ak}}
\def\rc{r_{\rm c}}
\def\rcu{r_{\rm c1}}
\def\rcd{r_{\rm c2}}
\def\rh{r_0}
\def\rs{r_1}
\def\rhoh{\rho_0}
\def\rhos{\rho_1}
\def\rhon{\rho_{\rm N}}
\def\rhou{\rho_{\gau}}
\def\rhod{\rho_{\gad}}
\def\roi{\rho _{\rm k}}
\def\roqi{\varrho _{\rm k}}

\def\rhotils{\tilde\rhos}
\def\rhotilh{\tilde\rhoh}
\def\rotilq{\tilde\varrho}
\def\sa{s_{\rm a}}

\def\sacp{s_{\rm ac}^+}
\def\sacm{s_{\rm ac}^-}
\def\sM{s_{\rm M}}
\def\sigr{\sigma_{\rm r}}
\def\sigt{\sigma_{\rm t}}
\def\un{U_{\rm N}}
\def\us{U_1}
\def\uh{U_0}
\def\uss{U_{11}}
\def\ush{U_{10}}
\def\uhs{U_{01}}
\def\uhh{U_{00}}
\def\wsh{W_{10}} 
\def\whs{W_{01}} 
   
\documentstyle[11pt,aaspp4]{article}

\slugcomment{In press on ApJ, Main Journal}

\lefthead{Ciotti L.}
\righthead{Two--Component Galaxy Models: Phase--Space Constraints}

\begin{document}

\title{Modelling Elliptical Galaxies:\\
       Phase--Space Constraints on\\
       Two--Component ($\gau$,$\gad$) Models}

\author{L. Ciotti$^{1,2}$}
\affil{$^1$Osservatorio Astronomico di Bologna, via Ranzani 1, 40127 Bologna, 
       Italy}
\affil{$^2$Scuola Normale Superiore, Piazza dei Cavalieri 7, 56126 Pisa,
       Italy}


\begin{abstract}
In the context of the study of the properties of the mutual mass
distribution of the bright and dark matter in elliptical galaxies, I
present a family of two--component, spherical, self--consistent galaxy
models, where one density distribution follows a $\gau$ profile, and
the other a $\gad$ profile [hereafter $(\gau,\gad)$ models], with
different total masses and ``core'' radii.  A variable amount of
(radial) orbital anisotropy is allowed in both components, following
the Osipkov--Merritt parameterization. For these models, I derive
analytically the necessary and sufficient conditions that the model
parameters must satisfy in order to correspond to a physical system
(the so--called model consistency).  Moreover, the possibility of
adding a black hole at the center of radially anisotropic $\gamma$
models is discussed, determining analytically a lower limit of the
anisotropy radius as a function of $\gamma$.  The analytical
phase--space distribution function for $(1,0)$ models is presented,
together with the solution of the Jeans equations and the quantities
entering the scalar virial theorem.  It is proved that a globally
isotropic $\gamma=1$ component is consistent for any mass and core
radius of the superimposed $\gamma=0$ model; on the contrary, only a
maximum value of the core radius is allowed for the $\gamma=0$ model
when a $\gamma=1$ density distribution is added.  The combined effects
of mass concentration and orbital anisotropy are investigated, and an
interesting behavior of the distribution function of the anisotropic
$\gamma=0$ component is found: there exists a region in the parameter
space where a sufficient amount of anisotropy results in a consistent
model, while the structurally identical but isotropic model would be
inconsistent.
\end{abstract}




%

\section{Introduction}
When studying a stellar dynamical model (single or multi--component),
the fact that the Jeans equations have a physically acceptable
solution is not a sufficient criterion for the validity of the model:
the minimal requirement to be met by a physically acceptable model is
the positivity of the phase--space distribution function (DF) of each
distinct component.  A model satisfying this essential requirement
(which is much weaker than the stability of the model) is called a
{\it consistent} model; moreover, when the total potential is
determined by the model density components through the Poisson
equation, the model is called {\it self--consistent}.  In other words,
a self--consistent model represents a physically acceptable,
self--gravitating system.

Two general strategies can be used to construct a (self) consistent
model, or check whether a proposed model is (self) consistent: they are
commonly referred to as the ``$f$--to--$\rho$'' and the
``$\rho$--to--$f$'' approaches (\cite{bt87}, Chap. 4, hereafter
BT87). An example of the first approach is the extensive survey of
self--consistent two--component spherical galaxy models carried out by
Bertin and co--workers (Bertin, Saglia, and Stiavelli 1992), where
they assume for the stellar and dark matter components two
distribution functions of the $f_{\infty}$ form (and so positive by
choice; \cite{bs84}). The main problem with this approach is that
generally the spatial density is not expressible in terms of simple or
at least well known functions, and so only numerical investigations
are usually feasible.

In the second approach, the density distribution is given, and
assumptions about the model internal dynamics are made, making the
comparison with the data simpler. The difficulties inherent in the
operation of recovering the DF in many cases prevent a simple
consistency analysis. In particular, in order to recover the DF of
spherical models with anisotropy, the Osipkov--Merritt technique
(\cite{osi79,mer85a}, hereafter OM) has been developed from the
original Eddington (1916) method for isotropic systems, and widely
used. Examples of {\it numerical} application of the OM inversion to
one and two--component spherical galaxies can be found in the
literature (see, e.g., \cite{cp92}, hereafter CP92;
\cite{hio94}; Carollo, de Zeeuw, and van der Marel 1995a; \cite{cl97}, 
hereafter CL97).  If one is just interested in the (self) consistency
of a stellar system the previous methods obviously give "too much",
i.e., give the full DF.  In the OM framework, a simpler approach in
order to check the (self) consistency of spherically symmetric,
multi--component models, is given by the method described in CP92.
This method uses directly the radial density profile of each component
and of the total potential, and gives necessary and sufficient
conditions for the model (self) consistency, avoiding the necessity of
recovering the DF itself.  Moreover, since it requires only spatial
differentiation and inequality checks, this method is best suited for
analytical investigations.

The importance of studying multi--component galaxy models cannot be
underestimated: in fact it is now accepted that a fraction of the mass
in galaxies and clusters of galaxies is made of a dark component,
whose density distribution -- albeit not well constrained by
observations -- differs from that of the visible one (see, e.g.,
\cite{bbb94,czm95b,bc97,gjsb98}).  Moreover, there is  
increasing evidence for the presence of massive black holes (BHs) at the
center of most (if not all) elliptical galaxies (see, e.g.,
\cite{hft94,vzr97a}, van der Marel, de Zeeuw, and Rix 1997b,
\cite{r98}).  It follows that the obvious generalization of the
one--component spherical models (the dynamicists zero--th order
approximation of real galaxies) is not only in the direction of the
actively developed modeling of axisymmetric and triaxial systems (see,
e.g., de Zeeuw (1996) for a review) but also in the direction of the
construction of two--component analytical models, and in the study of
their phase--space properties, a field far less developed. From this
point of view the $1^{\rm st}$ order approximation of real galaxies is
the construction of analytical, spherically symmetric, (self)
consistent {\it two--component} galaxy models.

Unfortunately, few examples of two--component systems in which both
the spatial density and the DF are analytically known are available,
namely the very remarkable axisymmetric Binney--Evans model
(\cite{bin81,eva93}), and the spherically symmetric two--component
Hernquist model (HH models, Ciotti 1996, hereafter C96), and so it
would be particularly interesting to find other members of this
exclusive club.  In C96 the (successful) choice of the Hernquist
density distribution (\cite{her90}, hereafter H90) as building block
for a two--component model with an analytical DF, was suggested by the
extremely simple (and algebraic) expression of its potential as a
function of radius. Moreover, the application of the CP92 method to HH
models revealed itself as both simple and fruitful, explaining many
properties of the derived DF.  In this line, a natural and promising
extension of the HH models is obtained by considering the wider family
of spherically symmetric, two--component, anisotropic $(\gau,\gad)$
models.  This family of models is made by the superposition of two
$\gamma$ models [see equation (7)] with different total masses,
scale--lengths, and slopes $\gamma$.  The mass concentration and
amount of mass of the two distributions are described by four free
parameters, and orbital anisotropy is allowed in both components,
following the OM prescription.  Because the Hernquist profile is
obtained from $\gamma=1$, it will referred as a $\gamma=1$ model, and
so with the adopted nomenclature the HH models discussed in C96 will
be called (1,1) models.  Note that an increasing interest in
$(\gau,\gad)$ models is seen in simulation and observational works:
e.g., Pellegrini and Ciotti (1998) used $(2,1)$ models in their
numerical simulations of hot gas flows in X--ray emitting elliptical
galaxies, Loewenstein and White (1999) used galaxy models similar to
$(1,1)$ models in the inner regions in order to observationally
constraint the properties of dark matter halos around ellipticals.

As expected, it is not possible to find the analytical DF for
$(\gau,\gad)$ models in the general case, but the fact that the DF of
(1,0) models with OM anisotropy is completely expressible in
analytical way, as proved here, is still of great interest.  The study
of (1,0) models is also useful for many other reasons: to provide an
analytical DF for a two--component system for which the virial
quantities and the analytical solution of the Jeans equation can be
found explicitly; to arrange initial conditions for numerical
simulations of two--component systems; to investigate the role of
anisotropy and mass distribution of each component in determining the
positivity of their DF. The availability of the analytical DF for a
two--component stellar system allows us to arrange with great accuracy
the initial conditions for numerical simulations aimed at
investigating the stability of galaxy models in the presence of dark
matter halos, or with a central BH. A work on stability of (1,0)
models is in progress (Londrillo and Ciotti, in preparation).

Strictly related to the last point above, is the trend shown in the
numerical investigations of two--component models described in CP92,
i.e., the difficulty of consistently superimposing a centrally peaked
distribution to a centrally flat one.  CP92 showed
numerically that King (1972) or quasi--isothermal density profiles can
not be coupled to a de Vaucouleurs (1948) model, because their DFs run
into negative values near the model center. On the contrary, the DF of
the de Vaucouleurs component is qualitatively unaffected by the
presence of centrally flat halos.  From this point of view, the C96
work on (1,1) models is complementary to the investigation of CP92: in
the (1,1) models the two density components are both centrally peaked,
and their DF is positive (in the isotropic case) for all the possible
choices of halo and galaxy masses and concentrations.

The implications of these findings have been not explored
sufficiently.  One could speculate that in the presence of a centrally
peaked dark matter halo, King--like elliptical galaxies should be
relatively rare, or that a galaxy with a central power--law density
profile cannot have a dark halo too flat in the center. In fact,
observations of the central surface brightness profiles of elliptical
galaxies (see, e.g.,
\cite {fer94,jaf94}; M{\o}ller, Stiavelli, M., and Zeilinger 1995, 
\cite{lau95,kor95,byu96}), and bulges of spirals (\cite{cs98}),
as well as high--resolution numerical simulations of dark matter halo
formation (see, e.g., \cite{dc91,white96}; Navarro, Frenk, and White
1997) seem to point in this direction.

In this paper, I further explore the trend that emerged in CP92 and in
C96, determining the limits imposed by phase--space constraints (i.e.,
the DF positivity) on the parameters describing the $(\gau,\gad)$
models and $\gamma$ models with a central BH [hereafter ($\gamma$,BH)
models]. I focus on the (1,0) models, in which one component is
centrally peaked ($\gamma=1$ density profile), and the other has a
more flat core ($\gamma=0$ density profile). With the aid of the
derived, analytical DFs, more stringent conclusions are then reached.

In Section 2, I briefly review the technique developed in CP92 and
applied in C96 to (1,1) models, and I formulate it in a way suitable
for its application to the present problem. In Section 3, the
$(\gau,\gad)$ models are introduced, as well as the CP92 method used
to discuss the limits imposed on their parameters by requiring the
positivity of the DF of the two components.  In Section 4, the DF for
the two components of (1,0) models are derived explicitly, and in
Section 5, the exact boundaries of the region of consistency in the
parameter space are obtained and compared to those given in Section
3. Finally, in Section 6, the main results are summarized. In the
Appendix, the analytical expressions for the velocity dispersion
profiles of both the (1,0) model components and the virial quantities
useful in applications, are derived in the general OM case.

\section{The Consistency of Multi--Component Systems}

As outlined in the Introduction, a stellar system described as a sum
of different density components $\roi$ is called consistent if {\it
each} $\fid$ is non--negative over all the accessible phase--space;  a
consistent, self--gravitating system is called self--consistent.  The
technique developed in CP92 permits us to check whether the DF of a
multi--component spherical system, where the orbital anisotropy of
each component is modeled according to the OM parameterization, is
positive, {\it without} calculating it effectively. In the OM
formulation, the radially anisotropic case is obtained as a consequence
of assuming $f=f(Q)$ with:
\begin{equation}
Q=\eps-{L^2\over 2\ra^2},
\end{equation}
where $\eps$ and $L$ are respectively the relative energy and the
angular momentum modulus per unit mass, $\ra$ is the so--called {\it
anisotropy radius}, and $f(Q)=0$ for $Q\leq 0$. With this assumption,
the models are characterized by radial anisotropy increasing with
galactic radius, and in the limit as $\ra\to\infty$, the velocity
dispersion tensor becomes globally isotropic. For a multi--component
spherical system, the simple relation between energy and angular
momentum prescribed by equation (1) allows us to express the DF of the
k--th component as:
\begin{equation}
\fid (\qi)={1\over\sqrt{8}\pi^2}{d\over d\qi}\int_0^{\qi} 
{d\roqi\over d\psit}{d\psit\over\sqrt{\qi-\psit}}=
{1\over\sqrt{8}\pi^2}\int _0^{\qi} 
{d^2\roqi\over d\psit^2}{d\psit\over\sqrt{\qi-\psit}},
\end{equation}
where 
\begin{equation}
\roqi (r)=\left (1+{r^2\over\rai^2}\right) \roi (r),
\end{equation}
$\psit (r)=\sum_k\psii (r)$ is the total relative potential, $\qi
=\eps-L^2/2\rai^2$, and $0\leq\qi\leq\psit (0)$. The second
equivalence in equation (2) holds for untruncated systems with finite
total mass, as the models discussed here (see, e.g., BT87, p.240). In
C96, the original CP92 technique was applied to (1,1) models using the
relative potential $\psii$ of the investigated component as the
independent variable, but here the radius $r$ is found to be a more
convenient variable. So, we have now
\medskip
\parn
{\bf Theorem}: 
A {\it necessary condition} (NC) for the non--negativity of $\fid$ is:
\begin{equation}
{d\roqi(r)\over dr}\leq 0,\quad 0\leq r \leq\infty .
\end{equation} 
If the NC is satisfied, a {\it strong sufficient condition} 
(SSC) for the non--negativity of $\fid$ is:
\begin{equation}
{d\over dr}\left[{d\roqi(r) \over dr}
{r^2\sqrt {\psit(r)}\over\mt (r)}\right]\geq 0,
\quad 0\leq r\leq\infty .
\end{equation}
Finally, a {\it weak sufficient condition} (WSC\footnote{The WSC is
better suited than the SSC for analytical investigations,
due to the absence of the weighting square root of the total
potential.})  for the non negativity of $\fid$ is:
\begin{equation}
{d\over dr}\left[
{d\roqi(r) \over dr}{r^2\over\mt (r)}\right]\geq 0,
\quad 0\leq r\leq\infty .
\end{equation}
\parn
{\bf Proof}: See CP92 and C96.
\parn
Some considerations follow from looking at the above conditions. The
first is that the violation of the NC [equation (4)] is connected only
to the radial behavior of $\roi$ and the value of $\rai$, and so this
condition applies independently of any other component added to the
model. Obviously, the condition imposed by equation (4) is only
necessary, so $\fid$ can be negative (and so the k component will be
inconsistent) even for values of model parameters allowed by the
NC. This is due to the radial behavior of the integrand in equation
(2), which not only depends on the particular $\roi$ and $\rai$, but
also on on the total potential.  To summarize: a model failing the NC
is {\it certainly inconsistent}, a model satisfying the NC {\it can be
consistent}. The second consideration is that a model satisfying the
WSC (or the more restrictive SSC) is {\it certainly} consistent, a
model failing the WSC (SSC) {\it can be consistent}, due to the
sufficiency of the conditions given by equations (5)-(6).  As a
consequence, the consistency of a model satisfying the NC and failing
the WSC (or the SSC) can be proved only by direct inspection of its
DF.  For example, if one find that for $\rai\leq\rai$(NC) the model is
inconsistent [i.e., equation (4) is not verified], while for
$\rai\geq\rai$(WSC)$\geq\rai$(SSC) the model is consistent [i.e.,
equations (5) or (6) are verified], this means that the true critical
anisotropy radius ($r_{\rm akc}$) for that model must satisfy the
relation $\rai$(NC)$\leq r_{\rm akc}\leq\rai$(WSC)$\leq\rai$(SSC).  In
this case, the limitations on $\rai$ obtained from WSC or SSC, are
upper bounds on the lower limit $r_{\rm akc}$ for consistency.
Obviously, the situation can be more complicated [see equations
(38)-(39), and the following discussion].

In the next section, after presenting the $(\gau,\gad)$ models, the
analytical constraints on $\ra$ for the consistency of the
one--component OM anisotropic $\gamma$ models are derived, together
with a general result on the consistency of two--component isotropic
$(\gau,\gad)$ models.  Successively, more specific results for the
isotropic (1,0) models are proved.  Finally, a limitation on $\ra$ for
the OM anisotropic ($\gamma$,BH) models, is explicitly derived as a
function of $\gamma$.

\section {The $(\gau,\gad)$ Models}

Both density distributions of the $(\gau,\gad)$ models belong to
the widely explored family of the so--called $\gamma$ models
(\cite{deh93}, hereafter D93; \cite{car93,tre94}):
\begin{equation}
\rho (r)={3-\gamma\over 4\pi}
         {M\,\rc\over r^{\gamma}(\rc+r)^{4-\gamma}},\quad\quad
         M(r)=M\times\left({r\over \rc+r}\right)^{3-\gamma},\quad\quad
0\leq\gamma <3,
\end{equation}
where $M$ is the total mass and $\rc$ a characteristic
scale--length. The corresponding relative potential is given by
\begin{equation}
\Psi(r)={GM\over\rc (2-\gamma)}
        \left [1-\left({r\over r+\rc}\right)^{2-\gamma}\right],\quad
\Psi(r)={GM\over\rc}\ln{r+\rc\over r},
\end{equation}
where the second expression holds for $\gamma=2$. In the following,
the mass $M=\Mu$ and the characteristic scale--length $\rc=\rcu$ of
the $\gau$ model will be adopted as normalization constants, so that
from equation (7) it follows $\rhou(r)=\rhon\tilde\rhou(s)$ and
$\rhod(r)=\rhon\mu\tilde\rhod(s,\beta)$, where $s=r/\rcu$,
$\rhon=\Mu/\rcu^3$, $\mu=\Md/\Mu$, and $\rcd=\beta\rcu$.  The
fundamental ingredient in recovering the DF is the total potential
$\psit=\psiu+\psid$, where from equation (8)
$\psiu(r)=\psin\tilde\psiu(s)$ and
$\psid(r)=\psin\mu\tilde\psid(s,\beta)$, and $\psin=G\Mu/\rcu$.  With
this choice, the $(\gau,\gad)$ models are structurally determined by
fixing the four independent parameters $(\Mu,\rcu,\mu,\beta)$, with
the obvious condition $\mu\geq 0$ and $\beta\geq 0$.

For future reference, I give here the explicit expressions of the
density and the potential for the (1,0) models, for which
$(\Mu,\rcu,\rhou,\psiu)=(\ms,\rs,\rhos,\psis)$, and
$(\Md,\rcd,\rhod,\psid)=(\mh,\rh,\rhoh,\psih)$:
\begin{equation}
\rhos (r) =\rhon\rhotils (s)={\rhon\over 2\pi}{1\over s(1+s)^3},
\end{equation}
and
\begin{equation}
\rhoh (r) =\rhon\mu\rhotilh (s)=
           \mu{3\rhon\over 4\pi}{\beta\over (\beta+s)^4},
\end{equation}
where $s=r/\rs$, $\rhon=\ms/\rs^3$, 
$\mh=\mu\ms$, and $\rh=\beta\rs$; moreover,
\begin{equation}
\psis (r) =\psin\psitils (s)={\psin\over 1+s},
\end{equation}
\begin{equation}
\psih (r) =\psin\mu\psitilh(s)=
\psin \mu {\beta+2s\over 2(\beta+s)^2},
\end{equation}
with $\psin=G\ms/\rs$. 

\subsection{The Necessary and Sufficient Conditions for the $\gamma$ Models}

Here I study first the NC for the general case of the anisotropic
one--component $\gamma$ models, in order to determine analytically a
{\it critical} anisotropy radius such that a higher degree of radial
OM anisotropy (i.e., a smaller anisotropy radius) produces a negative
DF for some permitted value of $Q$, no matter what kind of halo
density distribution is added. The unit mass and unit length are the
total mass $M$ and the scale--length $\rc$ of the model, with
$\sa=\ra/\rc$.

As shown in Appendix A [equations (A1)-(A2)], for $2\leq\gamma <3$ the
NC is satisfied for $\sa\geq 0$, i.e., the possibility that $\gamma$
models with $\gamma\geq 2$ are assembled using only radial orbits is
left open by the NC.  On the contrary, for $0\leq \gamma <2$ the NC
requires
\begin{equation}
\sa\geq \sM\sqrt{{2-\gamma-2\sM\over\gamma+4\sM }},
\end{equation}
where $\sM=\sM(\gamma)$ is given by equation (A2).  In this case the
NC {\it proves} that $\gamma$ models with $0\leq \gamma <2$ cannot
sustain radial orbits only. In Fig. 1 the lower bound for the
anisotropy radius as a function of $\gamma$ derived from the NC is
shown. From the discussion in Section 2, it follows that all $\gamma$
models (one or multi--component) in the nearly triangular region under
the solid curve are inconsistent.

The WSC can be treated analytically for the one--component
$\gamma$ models, as shown in Appendix A [equations (A3)-(A6)],
and we obtain the following limitats on $\sa$:
\begin{equation}
\sa\geq \sM^{3/2}\sqrt{{3-\gamma-\sM\over 6\sM^2+2(1+\gamma)\sM+\gamma}},
\end{equation}
where $\sM=\sM (\gamma)$ is given by equations (A4)-(A6).  As
discussed in Section 2, the r.h.s. of equation (14) (represented in
Fig. 1 by the dotted line), is an upper limit on the lower bound for
the critical anisotropy radius as a function of $\gamma$: all
one--component $\gamma$ models in the region above the dotted line are
consistent.

\placefigure{fig1}

A stronger limitation on $\ra$ is obtained using the SSC, but
unfortunately this condition for a generic $\gamma$ results in a
trascendental equation that cannot be solved explicitly. However, as
shown in Appendix A, for the three values $\gamma=(0,1,3)$ the
solution can be derived explicitly.  For $\gamma=0$,
\begin{equation}
\sa\geq\sM\sqrt{{3(1+2\sM -\sM^2)\over 14\sM^2+10\sM+2}}\simeq 0.501,
\end{equation}
where $\sM =\sM(0)$ is given by equation (A8).  For $\gamma=1$,
equation (A9) shows that
\begin{equation}
\sa\geq\sM^{3/2}\sqrt{{3(3-2\sM )\over 28\sM^2+17\sM +4}}\simeq 0.250,
\end{equation}
where $\sM=\sM(1)$ is given by equations (A10)-(A11).  The numerical
application of the SSC to the $\gamma=2$ model (\cite{jaf83}) gives
$\sa\gsim 0.047$. Finally, the case $\gamma=3$ is trivial, the SSC
reduces to $\sa\geq 0$.  These four values are represented in Fig. 1
by black dots: all one--component $\gamma$ models in the region above
the dashed line are consistent.  Note how the SSC improves the
estimate of the lower bound of $\ra$ with respect to the WSC.

As already pointed out in Section 2, the true critical value of $\sa$
for any one--component $\gamma$ model is between the NC and the SSC
curves. For the $\gamma=0$ and $\gamma=1$ models these values,
determined directly from their DFs (Sections B1-B2 in Appendix B), are
$\simeq 0.445$ and $\simeq 0.202$, respectively.  Merritt (1985b)
derived the analytical DF for a totally radial Jaffe model (the
$\gamma=2$ case): its positivity implies that in this case the true
lower limit on $\sa$ is zero. In Fig. 1 these three values are
represented by black squares, and in Table 1 all the previous results
for the specific cases $\gamma= (0,1,2,3)$ are summarized: note how
the more the model is concentrated, the more radial anisotropy can be
supported. The true critical $\ra$ value for the OM anisotropic
one--component $\gamma$ models as a function of $\gamma$ is
numerically known (see Fig. 1 in
\cite{czm95a}).

\placetable{tbl-1}

\subsection{Sufficient Conditions for Isotropic $(\gau,\gad)$ Models}

In order to proceed further with this analytical discussion, and to
allow for the presence of a ``halo'' component, we will use the WSC
rather than the more complicated SSC. The following three results are
proven analytically in this Section:

\begin{enumerate}
\item In the case of globally isotropic two--component $(\gau,\gad)$ models
with $1\leq \gau <3$ and $0\leq
\gad\leq\gau$, the DF of the more peaked component $\gau$ is positive
over all the phase space, for all values of
$(\mu,\beta)=(\Mu/\Md,\rcd/\rcu)$. As a consequence, the $\gamma=1$ 
component of (1,0) models is consistent for all values of the
parameters $(\mu ,\beta)$.

\item In the case of globally isotropic (1,0) 
models, the WSC applied to the $\gamma=0$ density distribution
suggests the existence of a lower limit of $\mu=\mh/\ms$ as a function
of $\beta=\rh/\rs$.  In particular, for $\beta\leq 5/2$, i.e., when
the $\gamma=0$ component is sufficiently concentrated, all values of
$\mu$ can be accepted.  Using the analytical DF the existence of this
lower limit will be proved in Section 5.

\item In the case of anisotropic $\gamma$ 
models with a BH at their center, it is possible to determine
analytically a lower limit on $\ra$ as a function of $\gamma$ using
the WSC.
\end{enumerate}

The proof of the first result is conceptually straightforward but
algebraically cumbersome. In Appendix A [equations (A12)-(A13)] it is
proved that under the hypothesis assumed in point 1 above, equation
(6) is verified for all choices of $(\mu,\beta)$. I note explicitly
that this result contains as a particular case the fact -- already
proved in C96 -- that globally isotropic (1,1) models can be
self--consistently assembled for any choice of $(\mu,\beta)$. From
this general result it also follows that -- e.g. -- the same is true
for isotropic Jaffe+Jaffe models ($\gau=\gad=2$).  Finally,
considering that for $\rc\to 0$ the potential of $\gamma$ models
becomes that of a point mass [see equation (8)], the previous result
means that a BH of any mass can be added at the center of a globally
isotropic $\gamma$ model when $1\leq\gamma <3$. A different analysis,
based on a series expansion of the integral representation of the DF
for isotropic $\gamma$ models, shows that the true limit on $\gamma$
in order to allow for the presence of a BH of any mass at the models
center, is $\gamma >1/2$ (\cite{tre94}).

The result stated in point 2 above is proved in Appendix A [equation
(A14)]. For equation (6) to be satisfied for the isotropic $\gamma=0$
component of (1,0) models, a sufficient condition for the consistency
of this component is
\begin{equation} 
\mu\geq (2\beta-5)\beta^2.
\end{equation}
This requirement can be interpreted in two different ways. The first
is that, having fixed the ratio $\beta=\rh/\rs$, only for sufficiently
high mass ratios $\mu=\mh/\ms$ can the $\gamma=0$ component ``dilute''
the effect of the central cusp of the $\gamma=1$ model on the total
potential, and be consistent.  More specifically, when $\beta<5/2$
(i.e., when the $\gamma=0$ density distribution is sufficiently
concentrated), even a vanishing mass $\mh$ ($\mu\to 0$) can be
accepted, while for large $\beta$ only very large $\mu$ are allowed.
From another point of view, equation (17) tells us that having fixed
$\mu$, $\beta$ cannot be arbitrarily large, but in some sense the
concentration of the $\gamma=0$ component must adapt to the density
distribution of the $\gamma=1$ component.  The effect of the
concentration is much more important than the amount of mass: in fact
$\beta\lsim(\mu/2)^{1/3}$. This means that even increasing
considerably the mass ratio, the maximum value of $\rh$ allowed for
the $\gamma=0$ model grows only like the {\it cube} root of $\mu$.
The limitation $\beta\leq 5/2$ is only a {\it sufficient} condition
for the consistency of a $\gamma=0$ model coupled with a dominant
$\gamma=1$: a larger critical value for $\beta$ is expected from
direct inspection of the DF when $\mu\to 0$ (see Section 5).

The result presented in point 3 above can be interpreted as an
extension to the radially anisotropic case of the analysis performed
by Tremaine et al. (1994), and is proved in Appendix A by showing that
the WSC applied to the anisotropic ($\gamma$,BH) models with
$1\leq\gamma <3$ can be analytically discussed in the special case of
a {\it dominant} BH, i.e., assuming in equation (6) $\mt =M_{\rm BH}$
(and so $\psit=GM_{\rm BH}/r$). Unfortunately, in the non--asymptotic
case, the equation to be discussed is transcendental, and no
analytical discussion can be carried out.  At first, the assumption of
a dominant BH could appear as a very rough approximation of reality,
but this is not true: the constraint derived can be used as a safe
limitation when constructing models containing a BH of a realistic
mass at their centre.  As shown in Appendix A, [equation (A15)], for
$1\leq\gamma <3$,
\begin{equation}
\sa\geq\sM
\sqrt{(3-\gamma)(\gamma-2)+4(3-\gamma)\sM-2\sM^2\over
      12\sM^2+8(\gamma-1)\sM+\gamma(\gamma-1)},
\end{equation}
where $\sM=\sM(\gamma)$ is obtained by solving a fourth degree
algebraic equation.  In Fig. 1, the long--dashed line represents the
lower bound for $\sa$ as determined by the previous equation, while
the explicit values for $\gamma=(1,2,3)$ are given in Table 1.  In
particular, the critical value for the $\gamma=1$ model was already
derived as a limiting case of a (non--asymptotic) formula given in C96
[equation (15) there].

\section{The DF of (1,0) Models}

We can now proceed to the explicit recovery of the DF of the (1,0)
models.  Just as for the density and the potential, it is also useful
for the DF to work with dimensionless functions; the two components of
the DF are of the form $f=\fn\ftil(\mu,\beta;\Qtil)$ with
$\fn=\rhon\psin^{-3/2}$ and $0\leq\Qtil
=Q/\psin\leq\psitils(0)+\mu\psitilh(0)$.  The easiest way to compute
each DF is to use the first of the identities in equation (2). For the
evaluation of the integral one would be tempted to obtain
$\varrho(\psit)$, eliminating the radius from the modified density and
the total potential: formally, this can be done, but the resulting
expression for the radial coordinate involves a quadratic
irrationality, that, after insertion in equations (3), (9) and (10),
produces an intractable expression.  Here I follow another approach:
instead of eliminating the radius, the integration variable is changed
from the total potential to the radius itself. This is equivalent to a
remapping of the domain of definition of each $f$, from the range of
variation of $\psit$ to the range of variation of $r$, and leads us to
introduce the dimensionless radius $\nu$ using equations (11)-(12):
\begin{equation} 
\Qtil={1\over 1+\nu}+\mu{\beta+2\nu\over2(\beta+\nu)^2}, 
      \quad 0\leq \nu\leq\infty.  
\end{equation} 
As shown in Appendix B [equations (B1)-(B2)], with this change of
variable and after normalization to the dimensional scales of the
Hernquist density distribution, the DF for the $\gamma=1$ and
$\gamma=0$ components can be formally written as:
\begin{equation}
f(Q) =    \fis(Q)+{\fan(Q)\over\sa^2}=
          {\fn\over\sqrt{8}\pi^2}\left({d\Qtil\over d\nu}\right)^{-1} 
          {d\over d\nu}\left[\Ftilis (\nu)+{\Ftilan (\nu)\over\sa^2}\right], 
          \quad\nu=\nu (\Qtil), 
\end{equation}
where $\sa =\ra/\rs$, and the subscripts refer to the isotropic and
anisotropic parts of the DF, respectively. Following the procedure,
$f(Q)$ results from the elimination of $\nu$ between equations
(19)-(20). In the general case, i.e., for any choice of
$(\mu,\beta,\sa)$, $f(Q)$ [and the so called {\it differential energy
distribution} for each component as well, (see, e.g., BT87, p.242)]
can be recovered analytically [see equation (B3) for a proof of this
fact].  So, it is shown that in addition to the (1,1) models, the
(1,0) models are a class of two--component stellar systems in which
both the spatial density distributions, the solution of the Jeans
equations (see Appendix C), and the phase--space distribution
functions can be explicitly found. Unfortunately, the DF of the (1,0)
models results in a combination of elliptic functions, even more
complicated than the DF of the (1,1) models, and this limits their
applicability to special problems in which the DF is required to be
known with arbitrary precision or to be formally manipulated.

Here I present only the DFs for the two density distributions obtained
under the assumption of a dominant ``halo'' component; I derive the DF
for a $\gamma=1$ model with a dominant $\gamma=0$ halo
($\mu\to\infty$), and for a $\gamma=0$ model with a dominant
$\gamma=1$ halo ($\mu\to 0$). Technically, this reduces to the
assumption that the total potential is the potential of the halo
component only.  Even though it is a limiting case, the study of {\it
halo--dominated} models is interesting for several different reasons:
1) the formulae -- expressible using elementary functions -- are much
simpler than in the general case, and can be studied very easily,
making clearer the effect of the halo component on the DF; 2) the
halo--dominated case is the one that differs most from the case of the
corresponding one--component model, and so the differences are better
evident; 3) all the intermediate cases fall between the one--component
model and the halo--dominated one.  A comparison with more realistic
values of the halo masses is postponed to Section 5.

In the following paragraphs, the two DFs will be compared with those
of the corresponding one--component $\gamma=1$ and $\gamma=0$ models,
and the exact phase--space constraints will be derived and compared
with those obtained using the NC, WSC, and SSC in Section 3.

\subsection{The $\gamma=1$ Model Plus a $\gamma=0$ Dominant Halo}

The explicit expression for the DF of the $\gamma=1$ model with an
arbitrary degree of OM orbital anisotropy immersed in a dominant
$\gamma=0$ halo is derived here. Formally, this case corresponds to
the assumption of $\mu\to\infty$ in the total potential, i.e., $\psit
= \psih$, and so in equations (19)-(20)
\begin{equation}
\Qtil=\mu{\beta+2\nu\over 2(\beta+\nu)^2},\quad
\left({d\Qtil\over d\nu}\right)^{-1}=-{(\beta+\nu)^3\over\mu\nu}.
\end{equation}
After differentiation inside the integral in equation (B2) with
$\rotilq$ given by equations (3) and (9), and after a partial
fraction decomposition of the rational part of the integrand, one
obtains:
\begin{equation}
\Ftilis(\nu)={\beta+\nu\over\pi\sqrt{2\mu}\sqrt{\beta+2\nu}}
[H^0_1+\beta H^0_2-H^1_1-(1+\beta)H^1_2-(1+2\beta)H^1_3-3(\beta-1)H^1_4],
\end{equation}
and 
\begin{equation}
\Ftilan(\nu)={\beta+\nu\over\pi\sqrt{2\mu}\sqrt{\beta+2\nu}}
[2H^1_2-(5-2\beta)H^1_3-3(\beta-1)H^1_4].
\end{equation}
The $H$ functions depend on $\beta$ and $\nu$, and are defined as
\begin{equation}
H^z_n(\xi) = {2\over (\nu+\lambda)^n}
             \int_0^{\infty}{dx\over\sqrt{1+x^2}(x^2+\xi)^n},
\end{equation}
where
\begin{equation}
\xi = {\nu+z\over\nu +\lambda}, \quad\quad {\rm and}\quad\quad 
\lambda = {\beta\nu\over\beta+2\nu}.
\end{equation}
When $\xi=1$ and $n\geq 1$
\begin{equation}
H^z_n(1)={\sqrt{\pi}\,\Gamma (n)\over\Gamma (n-1/2)(\nu+\lambda)^n},
\end{equation}
where $\Gamma$ is the complete gamma function.  When $\xi\neq
1$ the recursion formula
\begin{equation}
H^z_{n+1}(\xi)=-{1\over n(\nu+\lambda)}{dH^z_n(\xi)\over d\xi}=
{(-1)^n\over n!(\nu+\lambda)^n}{d^nH^z_1(\xi)\over d\xi ^n}
\end{equation}
holds, and so the explicit evaluation of $H^z_1(\xi)$ suffices:
\begin{equation}
H^z_1(\xi)={2\over\nu+\lambda}\cases{
           {\arccos\sqrt{\xi}\over\sqrt{\xi (1-\xi)}},
           &if $0\leq\xi <1$;\cr
           {\arccosh\sqrt{\xi}\over\sqrt{\xi (\xi-1)}},
           &if $\xi >1$.\cr
           }
\end{equation}
In order to distinguish between the two cases $\xi>1$ and $0\leq\xi
<1$, a careful discussion is needed. From equation (25) the value
$z=0$ corresponds to $\xi =\nu/(\nu+\lambda)<1\quad\forall
(\nu,\beta)$, and so the first of equations (28) must be used for the
evaluation of all $H^0_n$ functions.  More complicated is the case
$z=1$, when $\xi=(\nu+1)/(\nu+\lambda)$: note that for $\beta\geq 0$,
$\lambda$ is a monotonically increasing function of $\nu$, with
$\lambda=0$ for $\nu=0$ and $\lambda\to\beta /2$ for
$\nu\to\infty$. As a consequence, it follows that $\forall\nu$,
$0<\beta\leq 2\Rightarrow\xi>1$. When instead, $\beta>2$,
$\exists\,\nucr =\beta/(\beta-2)$ so that $\nu <\nucr\Rightarrow
\xi>1$, $\nu =\nucr\Rightarrow \xi =1$, and $\nu>\nucr\Rightarrow \xi
<1$.  This completes the derivation of the DF for the $\gamma=1$
component.

In Fig. 2 the comparison with the DF of the one--component $\gamma=1$
model (solid line), in case of global isotropy and for a specific
value of the anisotropy radius, is given. Such DF was given in H90 as
a function of $\Qtil$, but for consistency with the present work it is
derived in Appendix B as a function of $\nu$ [equations (B4)-(B6)].
The formulae derived in this paragraph have been tested for many
values of $\beta$ and $\ra$ using a code that numerically recovers the
DF for spherically symmetric multi--component galaxy models with OM
anisotropy; we obtained extremely good agreement; in all cases the
maximum differences between the analytical and numerical DFs are much
less than 1 per cent.

\placefigure{fig2}

In the upper panel of Fig. 2 the isotropic case is presented. Note how for
$\beta>1$ the DF is more peaked than for the one--component $\gamma=1$
model, and the opposite holds when $\beta <1$: this behaviour was
already found in the isotropic (1,1) models (C96, Fig. 2).  In the lower
panel the anisotropic case is shown when $\sa=0.26$, near the
consistency limit for the one--component $\gamma=1$ model (see \S
3.1).  The main effect of anisotropy, as already found for (1,1) models
and $R^{1/m}$ models (CL97), is the appearance in the DF of a
depression well outside the galaxy center. Decreasing the anisotropy
radius, the depression deepens, running finally into negative values
for a critical value of $\sa$ (dependent on $\beta$) and making the
model inconsistent. Again, as already found for (1,1) models, this effect
is stronger for smaller $\beta$ values, i.e., a very concentrated halo
makes the DF more sensitive to the effects of anisotropy, while the
opposite is true for halos more diffuse than the $\gamma=1$ density
distribution.

\subsection{The $\gamma=0$ Model Plus a $\gamma=1$ Dominant Halo}

The explicit expression for the DF of a $\gamma=0$ model with an
arbitrary degree of OM orbital anisotropy, immersed in a dominant
Hernquist halo, is derived here. Formally this case corresponds to the
assumption of $\mu\to 0$ in the total potential, i.e., $\psit =
\psis$, and so in equations (19)-(20)
\begin{equation}
\Qtil={1\over 1+\nu};\quad
\left({d\Qtil\over d\nu}\right)^{-1}=-(1+\nu)^2.
\end{equation}
After differentiation inside the integral in equation (B2) with
$\rotilq$ given by equations (3) and (10), and after a partial fraction
decomposition of the rational part of the integrand, one obtains:
\begin{equation}
\Ftilis(\nu)={3\mu\beta\sqrt{1+\nu}\over 4\pi}\,4 G_5,
\end{equation}
and 
\begin{equation}
\Ftilan(\nu)={3\mu\beta\sqrt{1+\nu}\over 4\pi}\,
             (4\beta^2 G_5-6\beta G_4+2G_3),
\end{equation}
where
\begin{equation}
G_n(\beta,\nu) = \int_{\nu}^{\infty}\sqrt{{s+1\over s-\nu}}
                 {ds\over (\beta+s)^n}.
\end{equation}
When $\beta=1$, and $n\geq 2$
\begin{equation}
G_n(1,\nu)={\sqrt{\pi}\,\Gamma (n-1)\over \Gamma(n-1/2) (1+\nu)^{n-1}}.
\end{equation}
When $\beta\neq 1$ and $n\geq 2$ the recursion formula 
\begin{equation}
G_{n+1}(\beta,\nu)=-{1\over n}{dG_n(\beta,\nu)\over d\beta}=
{(-1)^{n-1}\over n!}{d^{n-1}G_2(\beta,\nu)\over d\beta^{n-1}}
\end{equation}
holds, and so the explicit evaluation of $G_2$ suffices:
\begin{equation}
G_2(\beta,\nu)={1\over\beta+\nu}+{1+\nu\over (\beta+\nu)^{3/2}}\cases{
               {1\over\sqrt{1-\beta}}
               \arctan \sqrt{{1-\beta\over\beta +\nu}},
               &if $0\leq\beta <1$;\cr
               {1\over\sqrt{\beta-1}}
               \arctanh \sqrt{{\beta -1\over\beta +\nu}},
               &if $\beta >1$.\cr
               }
\end{equation}
In Fig. 3 the comparison with the DF of the one--component $\gamma=0$
model (solid line), in the case of global isotropy and for a specific
value of the anisotropy radius, is given.  Such DF was given in D93 as
a function of $\Qtil$, but for consistency with the present work it is
derived in Appendix B as a function of $\nu$ [equations (B7)-(B9)].
As in the previous case, the derived formulae have been successfully
tested for many values of $\beta$ and $\ra$ by comparison with the
numerically derived DFs, obtaining maximum differences less than 1 per
cent in any case.

\placefigure{fig3}

In the upper panel the isotropic case is presented.  Note how for
$\beta<1$ the DF is more peaked than for the one--component $\gamma=0$
model, and the opposite holds when $\beta<1\footnote{Having defined
$\beta=\rh/\rs$, at variance with what happened in \S 4.1, a more
diffuse $\gamma=1$ component corresponds to $\beta <1$, and
vice--versa.}$. This behavior is similar to that found in the previous
section.  In the lower panel the anisotropic case is shown when
$\ra=0.65\rh$, near the consistency limit for the one--component
$\gamma=0$ model (see \S 3.1).  As for the $\gamma=1$ component, the
main effect of anisotropy is the appearance in the DF of a depression
well outside the galaxy center, and again the depression becomes
deeper and deeper decreasing the anisotropy radius. Finally, as in the
previous case, the effect of anisotropy is found to be more important
when the halo concentration increases.

\section{Consistency of (1,0) Models}

We move now to comment on the main similarities and differences
between the DFs of the two components of the (1,0) models, especially
considering the role of concentration and orbital anisotropy in
determining their consistency. For simplicity the discussion is
restricted to the halo--dominated cases.


The first important point addressed by using the DFs, is the study of
the effect of halo concentration in determining the consistency of the
two (1,0) model components in the case of {\it global isotropy}. The
effect of the $\gamma=0$ halo concentration on the consistency of the
globally isotropic $\gamma=1$ component, can be derived by the direct
inspection of the DF and confirms the analytical prediction obtained
using the WSC in \S 3.1.  In fact, it is found that the globally
isotropic $\gamma=1$ component is consistent independent of the
concentration and total mass of the superimposed $\gamma=0$ halo: {\it
only anisotropic} $\gamma=1$ component in (1,0) models can be
unphysical due to the presence of the $\gamma=0$ density distribution.

For the globally isotropic $\gamma=0$ component with a dominant
$\gamma=1$ halo, the situation is more complicate, because, in
accordance with the analysis presented in \S 3.2, its DF may become
negative in case of a high concentration of the external $\gamma=1$
component.  In fact, the DF becomes negative for $\eps\to\psis (0)$
when $\beta\gsim 5.233$, a larger value than the more conservative one
(5/2) derived using the WSC.  A closer look at this behavior, and a
comparison with the qualitatively different behavior exhibited by the
DF of the $\gamma=1$ component, is particularly instructive.  In fact,
while the DF of the $\gamma=1$ density distribution diverges at high
(relative) energies both in the one--component and in the
halo--dominated cases (Fig. 2), the DF of the $\gamma=0$ model is
divergent for high energies in the one--component case, but {\it
finite} in the halo--dominated one (Fig. 3). Moreover, when increasing
the $\gamma=1$ halo concentration (i.e., increasing $\beta=\rh/\rs$),
the central value of the DF associated with the $\gamma=0$ density
profile decreases monotonically, and, for $\beta$ greater than the
before mentioned critical value, it becomes negative, revealing the
model inconsistency.  It must be stressed that a similar behavior was
found in the numerical investigation of consistency of King (1972) and
quasi--isothermal halos added to a de Vaucouleurs (1948) density
distribution, carried out by CP92.  Also, note how the decrease of the
central value of the DF for increasing halo concentration is
reminiscent of that found by C96 for (1,1) models, even if in that
case the transition was found to be more discontinuous: the DF of a
$\gamma=1$ component in (1,1) models remains divergent at the center
for all finite concentrations of the other $\gamma=1$ component, and
becomes exactly zero at the center only when the halo is reduced to a
central BH (see Fig. 2 in C96). The qualitative discussion above can
be put on more quantitative grounds. In fact, in the halo--dominated
case, the central value of the DF of the $\gamma=0$ component is
easily derived for a generic $\beta$ using the formulae given in \S
4.2:
\begin{equation}
\tilde\fis^0={3\mu\over 8\sqrt{2}\pi^3} \cases{
       -{3(64\beta^3-240\beta^2+280\beta-105)\over 
           32(1-\beta)^{5/2}\beta^{9/2}}\arctan\sqrt{{1-\beta\over\beta}}
       -{16\beta^3-328\beta^2+630\beta-315\over 
           32(1-\beta)^2\beta^4},
       &if $0<\beta <1$,\cr
        {64\over 5},
       &if $\beta =1$,\cr
       -{3(64\beta^3-240\beta^2+280\beta-105)\over 
           32(\beta-1)^{5/2}\beta^{9/2}}\arctanh\sqrt{{\beta-1\over\beta}}
       -{16\beta^3-328\beta^2+630\beta-315\over 
           32(\beta-1)^2\beta^4},
       &if $\beta >1$;\cr
                                      }
\end{equation}
and
\begin{equation}
\tilde\fan^0={3\mu\over 8\sqrt{2}\pi^3} \cases{
        {3(8\beta^2-12\beta+5)\over 
           32(1-\beta)^{5/2}\beta^{5/2}}\arctan\sqrt{{1-\beta\over\beta}}
       +{(4\beta-3)(2\beta-5)\over 
           32(1-\beta-)^2\beta^2},
       &if $0<\beta <1$,\cr
        {4\over 5},
       &if $\beta =1$,\cr
        {3(8\beta^2-12\beta+5)\over 
           32(\beta-1)^{5/2}\beta^{5/2}}\arctanh\sqrt{{\beta-1\over\beta}}
       +{(4\beta-3)(2\beta-5)\over 
         32(\beta-1)^2\beta^2},
       &if $\beta >1$.\cr
                                    }
\end{equation}
The limiting value for $\beta$ in the isotropic case is obtained by
solving numerically the equation $\tilde\fis^0=0$ for $\beta >1$.  In
Fig. 4 (where the high--concentration case corresponding to $\beta >1$
is shown), the decrease of $\tilde\fis^0$ when $\beta$ increases is
apparent.

\placefigure{fig4}


The second important point addressed by using the DFs, is the study of
the {\it combined} effect of orbital anisotropy and halo concentration
in determining the consistency of the two (1,0) model components.  We
cannot expect a simple behavior, because -- as should be clear from
the previous sections -- halo concentration and anisotropy affect the
DF in {\it different} regions of the phase--space, i.e., the high
energy regions of the DF are more sensitive to concentration effects,
while the OM orbital anisotropy acts mainly at intermediate energies.
The simplest way to summarize the results is to express the
consistency limitations in terms of the anisotropy radius of each
component as function of $\beta$, determining in the parameter space
$(\sa,\beta)$ the critical regions where the models are consistent.
This approach is particularly useful because independent of the
specific form of the density profile of the investigated model, the
positivity requirement for each DF of an OM multi--component system over
all the phase--space can be expressed in term of the anisotropy radius
as a function of the other model parameters, due to the simple
appearance of $\sa$ in equation (20).

In fact, let be $A_+$ the set defined by the property that $\fis>0$ 
$\forall\nu\in A_+$. Then, from equation (20), 
\begin{equation}
\sa\geq\sacm = \sqrt{\max\left\{0,{\rm sup}\left[-{\fan(\nu)\over 
                     \fis(\nu)}\right ]_{\nu\in A_+}\right\}},
\end{equation}
is a first condition to be satisfied. Obviously, when $\fis>0$ over
all the phase--space (the common situation), $A_+$ coincides with the
total range of variation for $\nu$, and equation (38) is also the {\it
only} condition to be checked for the model consistency.  In this case
equation (38) shows that there is {\it at most} a lower bound for the
anisotropy radius, $\sacm$. For example, this is the case for the
$\gamma=1$ component in (1,0) and (1,1) models, or for one--component
anisotropic $\gamma$ models.

When the set $A_-$ (complementary to $A_+$) is not empty, i.e., $\fis
<0$ over some region of phase--space, a second inequality, derived
from equation (20), must necessarily be verified:
\begin{equation}
\sa\leq\sacp = \sqrt{{\rm inf}\left[{\fan (\nu)\over 
                       |\fis (\nu)|}\right ]_{\nu\in A_-}}.
\end{equation}
A general consequence of equations (38)-(39) valid for {\it all} single 
or multi--component spherically symmetric, radially anisotropic OM
models, is that the allowed region for consistency in the anisotropy
space is given by $\sacm <\sa < \sacp$.  Moreover, if $\fan <0$
$\forall\nu\in A_-$, or $\sacp < \sacm$, then the proposed model is
inconsistent.

The quantitative trend of $\sacm$ for the $\gamma=1$ density
distribution with a dominant $\gamma=0$ halo is shown in Fig. 5a
(solid line): for a given $\beta$ all values of $\sa$ higher than the
critical curve are acceptable. Note how an increase in the halo
concentration (a decreasing $\beta$) produces an increase of $\sacm$,
i.e., a very concentrated halo makes the other component more
sensitive to anisotropy effects, a behavior qualitatively anticipated
in \S 4.1, and already found for (1,1) models (C96, Fig. 5).

A more complicated (and more interesting) case is presented by the
halo--dominated $\gamma=0$ model. In this case we already know that,
due to the halo concentration, even the isotropic case can be
inconsistent, i.e., $\fis <0$. This means that $\sacp$ must also be
considered.  The trend of $\sacm$ is shown in Fig. 5ab (dotted line):
as in the previous case an increase of the minimum anisotropy radius
corresponds to an increasing halo concentration (i.e., to an
increasing $\beta$).  As $\beta$ increases above the critical value
$\simeq 5.233$, $\fis$ becomes negative at the center, and the
isotropic $\gamma=0$ component becomes inconsistent: in Fig. 5a this
region is contained in the box at the top--right, that is enlarged in
Fig. 5b.  Here the dotted line is again $\sacm$, and the dashed line
represents $\sacp$: for $5.233\lsim\beta\lsim 6.15$ the inequality
$\sacm<\sacp$ holds, and according to equations (38)-(39) the region
between the two curves corresponds to consistent $\gamma=0$
components.  This is a quite counterintuitive example of the combined
effect of an external potential and anisotropy on the consistency of
an anisotropic galaxy model, where an otherwise inconsistent isotropic
model is made consistent by orbital anisotropy!  Finally, for $\beta
\gsim 6.15$ no physically acceptable $\gamma=0$ components are
possible, even considering the effect of anisotropy.

A question arises: how well does the asymptotic analysis obtained in
the limit of dominant halos compare to the more realistic cases of
halos with finite mass?  An answer can be obtained by inspection of
Fig. 5a, where the dashed lines represent the limits on the anisotropy
radius obtained when considering a halo ten times as massive as the
component investigated. Note that when the halo is {\it more
concentrated} than the considered density component [large $\beta$ in
case of $\gamma=0$ model (dotted line) and small $\beta$ for the
$\gamma=1$ model (solid line)], the curves corresponding to the
asymptotic analysis and the dashed ones are indistinguishable for any
practical application.  On the contrary, a small departure appears
when the halo scale--length is substantially larger than that of the
considered density component, with the dashed curves approaching the
critical value for the anisotropy radius corresponding to the
one--component model (the two black dots). This is an obvious
behavior, since for any {\it finite} value of the halo mass, its
gravitational effect becomes weaker and weaker for larger and larger
halo scale--length.

\placefigure{fig5}

\section{Conclusions}

In this paper, an extensive analytical investigation of the
phase--space of two--component spherical galaxy models made of the sum
of a Hernquist density distribution and a $\gamma=0$ model with
different physical scales, is carried out. Following the simple
Osipkov--Merritt parameterization, a variable amount of orbital
anisotropy is allowed in each component. For these models, other
important properties useful in applications -- the velocity dispersion
components and the various energy terms entering the scalar virial
theorem -- can be expressed analytically, and are given in Appendix C.
The main results can be summarized as follows:

\begin{enumerate}

\item The necessary and sufficient conditions that the model parameters 
must satisfy, in order to correspond to a (1,0) system for which the
two physically distinct components have a positive DF are analytically
derived using the method introduced in CP92.  Some conditions are
obtained for the wider class of two--component $(\gau,\gad)$ models
[of which the (1,0) models are a special case].  In particular, it is
shown that the DF of the $\gau$ component in isotropic $(\gau,\gad)$
models is nowhere negative, independent of the mass and concentration
of the $\gad$ component, whenever $1\leq\gau <3$ and
$0\leq\gad\leq\gau$. As a special application of this result, it
follows that a BH of any mass can be consistently added at the center
of any isotropic member of the $\gamma$ family of models, when
$1\leq\gamma <3$. Two important consequences follow. The first is that
the consistency of isotropic (1,1) [or (1,BH)] models proved in C96
using an ``ad hoc'' technique is not exceptional, but a common
property of a large class of two--component $\gamma$ models: for
example, also isotropic two--component Jaffe ($\gamma=2$) or Jaffe+BH
models can be safely assembled. The second is that in two--component
isotropic models, the component with the steeper central density
distribution is usually the most robust against inconsistency.

\item It is shown that an analytic estimate of a minimum value of $\sa$
for one--component $\gamma$ models with a massive (dominant) BH at
their center can be explicitly found. As expected, this minimum value
decreases for increasing $\gamma$.

\item It is shown that the analytic expression for the DF of (1,0) models 
with general OM anisotropy can be found in terms of elliptic
functions. The special cases in which each one of the two density
components are embedded in a dominant halo are also discussed: under
this assumption the DFs can be expressed using just elementary
functions, allowing a detailed analytical investigation.

\item The region of the parameter space in which (1,0) 
models are consistent is explored using the derived DFs: it is shown
that, unlike the $\gamma=1$ component, the $\gamma=0$ component
becomes inconsistent when the halo is sufficiently concentrated, even
in the isotropic case.  This is an explicit example (albeit not so
extreme) of the result found by CP92, that numerically proved the
impossibility of adding a King or a quasi--isothermal halo to a de
Vaucouleurs galaxy.  In such models, the (isotropic) de Vaucouleurs
galaxy was found instead consistent over all the parameter space.

\item The combined effect of halo concentration and orbital anisotropy is
finally investigated. The trend of the minimum value for the
anisotropy radius as a function of the halo concentration is
qualitatively similar in both components, and to that found for (1,1)
models in C96: a more diffuse halo allows a larger amount of
anisotropy. A qualitatively new behavior is found and explained by
investigating the DF of the $\gamma=0$ component in the
halo--dominated case for high halo concentrations. It is analytically
shown that there exists a small region in the parameter space where a
sufficient amount of {\it anisotropy} can compensate the inconsistency
produced by the halo concentration on the structurally analogous --
but isotropic -- case.

\end{enumerate}

As a final remark, it can be useful to point out some general trends
that emerge when comparing different one and two--component models
with OM anisotropy, as those investigated numerically in CP92 and
CL97, and analytically in C96 and in this paper. The first common
trend is that OM anisotropy produces a negative DF outside the galaxy
center, while the halo concentration affects mainly the DF at high
(relative) energies.  The second is that the possibility of sustaining
a strong degree of anisotropy is weakened by the presence of a very
concentrated halo. The third is that in two--component models, in
cases of very different density profiles in the central regions, the
component with the flatter density is the most ``delicate'' and can
easily found to be inconsistent: particular attention should be paid
in constructing such models.

\acknowledgments

I would like to thank Giuseppe Bertin, Laura Greggio, and Silvia
Pellegrini for helpful comments and discussions. The referee, Stephen
Levine, is especially thanked for his comments that greatly improved
the paper.  This work has been partially supported by contracts
ASI-95-RS-152, ASI-ARS-96-70, and MURST--Cofin98.

\appendix

\section{Consistency Requirements}

\subsection{The NC and WSC for One--Component $\gamma$ Models}

The NC [equation (4)] for the anisotropic (one--component) $\gamma$
models imposes a limitation on the anisotropy radius:
\begin{equation}
\sa^2\geq {s^2 (2-\gamma-2s)\over \gamma+4s},
\quad  0\leq s\leq\infty .
\end{equation}
This inequality is true for $\sa^2$ larger than or equal to the
maximum of the function on the r.h.s. of equation (A1).  For
$2\leq\gamma$ the r.h.s. is strictly negative, and so all values of
$\sa$ satisfy the NC. When $0\leq\gamma <2$ the maximum is reached at
\begin{equation}
\sM (\gamma) ={4-5\gamma+\sqrt{(4-\gamma)(4+7\gamma)}\over 16},
\end{equation}
that after substitution in equation (A1), gives equation (13). 

The WSC [equation (6)] applied to the anisotropic one--component
$\gamma$ models gives the following inequality:
\begin{equation}
\sa^2\geq {s^3 (3-\gamma-s)\over 6s^2 +2(1+\gamma)s +\gamma},
\quad 0\leq s \leq\infty .
\end{equation}
After differentiation, one is left with the discussion of a cubic
equation. Its discriminant is negative for $0\leq\gamma <
(\sqrt{73}-5)/8$ and positive for $(\sqrt{73}-5)/8 <\gamma <3$. In
the first case two of the three real solutions are negative, and the
maximum of the r.h.s. of equation (A3) is reached at
\begin{equation}
\sM (\gamma) =
         {1-\gamma\over 3}+{2\sqrt{4-\gamma}\over 3}\cos\left[
         {1\over 3}\arctan{\sqrt{(15-4\gamma)(3-\gamma)(3-5\gamma-4\gamma^2)}
         \over 11+11\gamma-4\gamma^2}\right].
\end{equation}
In the second case, discarding the two complex conjugates roots, and
defining
\begin{equation}
s_0=(4-\gamma)[11+11\gamma-4\gamma^2+
      \sqrt{(15-4\gamma)(3-\gamma)(4\gamma^2+5\gamma-3)}],
\end{equation}
the maximum is reached at
\begin{equation}
\sM (\gamma) = 
    {s_0^{1/3}\over 6}+{1-\gamma\over 3}+{2(4-\gamma)\over 3 s_0^{1/3}}.
\end{equation}
Finally, when $\gamma=(\sqrt{73}-5)/8\simeq 0.443$, $\sM=
(\sqrt{73}+3)/8\simeq 1.443$.

\subsection{The SSC for the $\gamma=0$ and $\gamma=1$ Models}

The SSC [equation (5)] applied to the anisotropic one--component
$\gamma=0$ model gives the following inequality:
\begin{equation}
\sa^2\geq {3s^2(1+2s-s^2)\over 14s^2+10s+2},\quad 0\leq s\leq\infty .
\end{equation}
The maximum of the r.h.s. of equation (A7) can be obtained explicitly
solving a fourth--degree equation. The numerical value of the only
physically acceptable solution is
\begin{equation}
\sM (0)\simeq 1.3149,
\end{equation}
that after back substitution in equation (A7) gives equation (15).

The SSC applied to the anisotropic one--component $\gamma=1$ model
gives the following inequality:
\begin{equation}
\sa^2\geq {3s^3(3-2s)\over 28s^2+17s+4},\quad 0\leq s\leq\infty.
\end{equation}
After differentiation, discarding the two complex conjugates roots of the 
resulting cubic equation, and defining 
\begin{equation}
s_0=681939+84\sqrt{35887965},
\end{equation}
the maximum of the r.h.s. of equation (A9) is reached at
\begin{equation}
\sM (1) ={s_0^{1/3}\over 168}-{3\over 56}+{1987\over 56s_0^{1/3}}
         \simeq 0.9116,
\end{equation}
that after back substitution in equation (A9) gives equation (16).

\subsection{The WSC for Isotropic $(\gau,\gad)$}

With the aid of the WSC, the positivity of the DF for the $\gau$
density distribution of the globally isotropic two--component
$(\gau,\gad)$ models, where $1\leq\gau <3$ and $0\leq\gad\leq\gau$, is
here discussed.  Having computed the derivatives prescribed by
equation (6), we have to investigate the positivity of a rational
expression, whose denominator is strictly positive $\forall (\gau,
\gad)$ and $\forall (s,\beta)$; the numerator factorizes in a strictly
positive function and in a transcendental expression.  Defining
$\gau=1+\epu$ (with $0\leq\epu <2$) and $\gad=\gau-\epd$ (with
$0\leq\epd\leq\gau$), the trascendental factor reads:
\begin{equation}
2(s+\beta)^4(1+1/s)^{\gau}[6s^2+2(1+\gau)s+\gau]+
\mu(s+1)^3(1+\beta/s)^{\gad}F(s,\beta,\gau,\gad),
\end{equation}
where the first addend is strictly positive, and
\begin{equation}
F(s,\beta,\gau,\gad)= 12s^3+4[(5-\epu+\epd)\beta+2\epu]s^2+
                      [(10+5\epu-\epu^2+5\epd +\epu\epd)\beta+\epu\gau]s+
                      \beta\gau(2+\epd).
\end{equation}
In the range of values for $\epu$ and $\epd$ the positivity of $F$ is
easy proved $\forall (s,\beta)\geq 0$, and so condition (6) is
verified $\forall (\mu,\beta)$.

The application of the WSC to the globally isotropic $\gamma=0$ component 
of a (1,0) model is simple, leading to the discussion of:
\begin{equation} 
\mu\geq {-(3s+5-2\beta)(s+\beta)^3\over 
       (3s+\beta)(s+1)^3},\quad 0\leq s\leq\infty.
\end{equation}
First, note that for $\beta <5/2$ the previous inequality is satisfied
$\forall\mu\geq 0$. When $\beta\geq 5/2$, the maximum of the r.h.s. of
equation (A14) over the interval $0\leq s\leq\infty$ is reached at
$s=0$, and so the condition given in equation (17) is derived.

\subsection{The WSC for Anisotropic ($\gamma$,BH) Models}

In this case we assume in equation (6) $\mt=M_{\rm BH}$, and, from the
previous discussion, we restrict to $\gamma$ models with $1\leq\gamma
<3$.  After the computation of the derivatives, we have to investigate
the following inequality
\begin{equation}
\sa^2\geq {s^2[(3-\gamma)(\gamma-2)+4(3-\gamma)s-2s^2]\over
12s^2+8(\gamma-1)s+\gamma(\gamma-1)},\quad 0\leq s\leq\infty.
\end{equation}
After differentiation, one is left with the discussion of a quartic
equation, and it can be shown that there exists only one maximum,
located at $\sM=\sM(\gamma)\geq 0$.  The explicit expression for
$\sM(\gamma)$ is not very useful, and so is not reported here. In any
case, it can be of interest the explicit values of $\sM(\gamma)$ for
$\gamma=(1,2,3)$. After easy computations, one finds
\begin{equation}
\sM(1)=2,\quad
\sM(2)={(54+6\sqrt{33})^{1/3}\over 6}+
       {2\over (54+6\sqrt{33})^{1/3}}\simeq 1.191,\quad
\sM(3)=0,
\end{equation}
that after substitution in equation (18) gives the values reported in
Table 1.

\section{The DF of (1,0) Models}

Here the main steps required for the determination of the DF of each
component of (1,0) models are briefly described. Changing the
integration variable from the total potential to the radius, and after
normalization to the physical scales of the $\gamma=1$ component,
equation (2) becomes:
\begin{equation}
f(Q)={\fn\over\sqrt{8}\pi^2}\left({d\Qtil\over d\nu}\right)^{-1}
{d\Ftil(\nu)\over d\nu},
\end{equation}
where the relation between $\nu$ and $\Qtil$ is given by equation (19), and
\begin{equation}
\Ftil(\nu)=-\int_{\nu}^{\infty}{d\tilde\varrho\over ds}
{ds\over\sqrt{\psitilt (\nu)-\psitilt (s)}}.
\end{equation}
Note that a negative sign appears in front of the previous integral,
due to the monotonic decrease of the relative potential with radius.
From equations (11)-(12)
\begin{equation}
{1\over\sqrt{\psitilt (\nu)-\psitilt(s)}}=
\sqrt{{1+s\over s-\nu}}
{\sqrt{2(1+\nu)}(\beta+\nu)(\beta+s)\over\sqrt{As^2+Bs+C}},
\end{equation}
where $A,B,C$ are constants depending on $(\mu,\beta,\nu)$. Finally,
observing that for each component $d\tilde\varrho/ds$ is a rational
function of $s$, with the standard change of the integration variable
$(1+s)/(s-\nu)=t^2$, it follows that the integral in equation (B2) can
be expressed in terms of elliptic integrals.

\subsection{The DF of the One--Component $\gamma=1$ 
            Model as a Function of $\nu$}

In this case, in equation (B1)
\begin{equation}
\Qtil={1\over 1+\nu};\quad
\left({d\Qtil\over d\nu}\right)^{-1}=-(1+\nu)^2.
\end{equation}
After differentiation under the integral in equation (B2), with
$\rotilq$ given by equations (3) and (7), and $\psitilt = \psitils$,
and after a partial fraction decomposition of the rational part of the
integrand, one obtains:
\begin{equation}
\Ftilis(\nu)=-{\sqrt{1+\nu}\over 2\pi}
             \left[
             {2\over 15}{15\nu^2+50\nu+59\over (1+\nu)^3}-{1\over\nu}-
             {1+\nu\over\nu^{3/2}}\arctan {1\over\sqrt{\nu}}
             \right],
\end{equation}
and 
\begin{equation}
\Ftilan(\nu)=-{\sqrt{1+\nu}\over 2\pi}{8\over 15}{1-5\nu\over (1+\nu)^3}.
\end{equation}
A check of the derived formulae is obtained by substitution of
$\nu=\nu(\Qtil)$ as given by inverting the first of equations (B4),
and comparing the result with the DF given in H90.

\subsection{The DF for the One--Component $\gamma=0$ 
            Model as a Function of $\nu$}

In this case, the scale--length $\rh$ of the $\gamma=0$ component and
its mass $\mh$, are assumed as physical scales, and so
$\fn=(\mh/\rh^3)(G\mh/\rh)^{-3/2}$. This is formally equivalent to
assume $\beta=1$ and $\mu=1$ in equations (10) and (12), and so in
equation (B1):
\begin{equation}
\Qtil ={1+2\nu\over 2(1+\nu)^2};\quad
\left({d\Qtil\over d\nu}\right)^{-1}=-{(1+\nu)^3\over\nu}.
\end{equation}
After differentiation under the integral in equation (B2) with
$\rotilq$ given by equations (3) and (9), and $\psitilt = \psitilh$,
and after a partial fraction decomposition of the rational part of the
integrand, one obtains:
\begin{equation}
\Ftilis(\nu)=-{3\sqrt{2}\over 4\pi}
              \left[
              {2\sqrt{1+2\nu}(15\nu^2+22\nu+11)\over 3(1+\nu)^3}-
              {4(5\nu^2+4\nu+2)\over (1+\nu)^2}\arctanh {1\over\sqrt{1+2\nu}}
              \right],
\end{equation}
and 
\begin{equation}
\Ftilan(\nu)=-{3\sqrt{2}\over 4\pi}
             \left[
             {\sqrt{1+2\nu}(9\nu^2+2\nu+1)\over 3(1+\nu)^3}-
             {6\nu^2\over (1+\nu)^2}\arctanh {1\over\sqrt{1+2\nu}}
             \right].
\end{equation}
A check of the derived formulae is obtained by substitution of
$\nu=\nu(\Qtil)$ as given by inverting the first of equations (B7),
and comparing the result with the DF given in D93\footnote{The
anisotropic part of the DF given in D93 should be multiplied by 2.}.

\section{The Velocity Dispersions and Virial Quantities}

Here I present the main dynamical quantities of the model discussed in
the paper.  The radial component $\sigr^2$ of velocity dispersion
$\sigma^2=\sigr^2+\sigt^2$ in the OM parameterization can be written
for each component as
\begin{equation} 
\rho(r)\sigr^2(r)={A(r)+\ra^2 I(r)\over r^2+\ra^2} 
\end{equation}
where
\begin{equation}
A(r)=G\int_r^{\infty}\rho(r)\mt(r)dr,\quad\quad
I(r)=G\int_r^{\infty}{\rho(r)\mt(r)\over r^2}dr,
\end{equation}
(\cite{bm82}); once obtained $\sigr^2$, the tangential
velocity dispersion is given by
\begin{equation}
\sigt^2(r)={2\ra^2\over r^2+\ra^2}\sigr^2(r).
\end{equation}
Other quantities of interest in applications as the global energies
entering the scalar virial theorem are derived in the next paragraphs.

\subsection{The $\gamma=1$ Component}

Due to the presence of the $\gamma=0$ component,
$\tilde\Is=\tilde\Iss+\mu\tilde\Ish$ and
$\tilde\As=\tilde\Ass+\mu\tilde\Ash$, where the dimensional
coefficients of the velocity dispersions and energies are $\psin$ and
$\un=\ms\psin$, respectively.  After normalization and integration one
finds
\begin{equation}
\tilde\Iss={1\over 2\pi}\ln {1+s\over s} -
           {12s^3+42s^2+52s+25\over 24\pi (1+s)^4},
           \quad\quad
\tilde\Ass={1+4s\over 24\pi (1+s)^4}.
\end{equation}
The interaction with the $\gamma=0$ halo is described by the two contributions
\begin{equation}
\tilde\Ish={3\over\pi (\beta-1)^5}\ln {\beta+s\over 1+s} 
          -{(2s+\beta+1)[6s^2+6s(\beta+1)-\beta^2+8\beta-1]\over
            4\pi (\beta-1)^4(1+s)^2(\beta+s)^2},
\end{equation}
and
$$
\tilde\Ash={(\beta^2+4\beta+1)\over 2\pi (\beta-1)^5}
           \ln {\beta+s\over 1+s} -
           \quad\quad\quad\quad\quad\quad
           \quad\quad\quad\quad\quad\quad
           \quad\quad\quad\quad\quad\quad
$$
\begin{equation}
           {2(\beta^2+4\beta+1)s^3+3(\beta+1)(\beta^2+4\beta+1)s^2+
            2\beta(5\beta^2+8\beta+5)s+6\beta^2(\beta+1)\over 
            4\pi (\beta-1)^4(\beta+s)^2(1+s)^2}.
\end{equation}
When $\beta=1$
\begin{equation}
\tilde\Ish={1\over 10\pi (1+s)^5};\quad
\tilde\Ash={10s^2+5s+1\over 60\pi (1+s)^5}.
\end{equation}

The gravitational energy of the $\gamma=1$ component is given by
$\tilde\us=\tilde\uss +\mu\tilde\ush$, where $\uss=-2\pi\int\rhos\psis
r^2dr$ is the contribution due to the self--interaction, and $\ush
=-4\pi\int\rhos\psih r^2dr$, is due to the $\gamma=0$ halo potential.
After normalization,
\begin{equation}
\tilde\uss=-{1\over 6};\quad
\tilde\ush=-{\beta^2-5\beta-2\over 2(\beta-1)^3}
           -{3\beta\ln\beta\over (\beta-1)^4},
\end{equation}
with $\tilde\ush=-1/4$ for $\beta=1$.  As well known, the scalar
virial theorem for a multi--component system reads $\tilde
K_*=(|\tilde\uss|+\mu|\tilde\wsh|)/2$, where the interaction energy is
$\wsh=4\pi\int r^3\rhos (d\psih/dr)dr$, and after normalization,
\begin{equation}
\tilde\wsh=-{\beta^2+10\beta+1\over (\beta-1)^4}
           +{6\beta(\beta+1)\ln\beta\over(\beta -1)^5},
\end{equation}
with $\tilde\wsh=-1/10$ for $\beta=1$.
 
\subsection{The $\gamma=0$ Component}

In this section the normalization constants are the physical scales of
the $\gamma=0$ component, i.e., $s=r/\rh$, $\mu=\ms/\mh$ and $\beta
=\rs/\rh$.  Consistently with this choice, now $\psin=G\mh/\rh$,
$\un=\mh\psin$, $\tilde\Ih=\tilde\Ihh+\mu\tilde\Ihs$, and
$\tilde\Ah=\tilde\Ahh+\mu\tilde\Ahs$.  After normalization and
integration one finds
\begin{equation}
\tilde\Ihh={1+6s\over 40\pi (1+s)^6},
           \quad\quad
\tilde\Ahh={20s^3+15s^2+6s+1\over 80\pi (1+s)^6}.
\end{equation}
The interaction with the $\gamma=1$ halo is described by the two contributions
\begin{equation}
\tilde\Ihs={3\over\pi (\beta-1)^5}\ln {1+s\over\beta+s}
    +{12s^3+6(\beta+5)s^2-2(\beta^2-8\beta-11)s+\beta^3-5\beta^2+13\beta+3
           \over 4\pi (\beta-1)^4(1+s)^3(\beta+s)},
\end{equation}
and
$$
\tilde\Ahs={3\beta (\beta+1)\over 2\pi (\beta-1)^5}
           \ln {1+s\over \beta+s} +
           \quad\quad\quad\quad\quad\quad
           \quad\quad\quad\quad\quad\quad
           \quad\quad\quad\quad\quad\quad
$$
\begin{equation}
           {6\beta(\beta+1)s^3+3\beta(\beta+5)(\beta+1)s^2+
           (3\beta^3+25\beta^2+7\beta+1)s+\beta(\beta^2+10\beta+1)
           \over 4\pi (\beta-1)^4(1+s)^3(\beta+s)}.
\end{equation}
When $\beta=1$
\begin{equation}
\tilde\Ihs={3\over 20\pi (1+s)^5};\quad 
\tilde\Ahs={10s^2+5s+1\over 40\pi (1+s)^5}.
\end{equation}

The gravitational energy of the $\gamma=0$ component is given by
$\tilde\uh=\tilde\uhh +\mu\tilde\uhs$, where $\uhh=-2\pi\int\rhoh\psih
r^2dr$ is the contribution due to the self--interaction, and $\uhs
=-4\pi\int\rhoh\psis r^2dr$, is due to the $\gamma=1$ halo potential.
After normalization,
\begin{equation}
\tilde\uhh=-{1\over 10};\quad
\tilde\uhs=-{2\beta^2+5\beta-1\over 2(\beta-1)^3}
           +{3\beta^2\ln\beta\over (\beta-1)^4},
\end{equation}
with $\tilde\uhs=-1/4$ for $\beta=1$, obviously equal to
$\tilde\ush$.  In this case the interaction energy is $\whs=4\pi\int
r^3\rhoh (d\psis/dr)dr$ and the virial theorem is $\tilde K_{\rm
h}=(|\tilde\uhh|+\mu|\tilde\whs|)/2$; after normalization one obtains
\begin{equation}
\tilde\whs={17\beta^2+8\beta-1\over 2(\beta-1)^4}
           -{3\beta^2(\beta+3)\ln\beta\over(\beta -1)^5}.
\end{equation}
For $\beta=1$, $\tilde\whs=-3/20$. 

\clearpage

\begin{deluxetable}{cccccc}
\footnotesize
\tablecaption{Critical values of $\sa=\ra/\rc$ for $\gamma$ models. The 
              last column refers to the case of a dominant central
              BH. 
              \label{tbl-1}}
\tablewidth{0pt}
\tablehead{
\colhead{$\gamma$} & \colhead{NC}   & \colhead{True\tablenotemark{a}}   & 
\colhead{SSC}      & \colhead{WSC}  & \colhead{WSC$_{\rm BH}$} 
} 
\startdata
0 & 0.354 & 0.445 & 0.501 & 0.556 & ----- \nl
1 & 0.128 & 0.202 & 0.250 & 0.310 & 0.707 \nl
2 & 0.000 & 0.000 & 0.047 & 0.107 & 0.309 \nl
3 & 0.000 & 0.000 & 0.000 & 0.000 & 0.000 \nl
 
\enddata
\tablenotetext{a}{Obtained from the model DF.}

\end{deluxetable}
\clearpage

\clearpage

\figcaption[]{The limits on the anisotropy radius for the consistency of 
the one--component $\gamma$ models as a function of $\gamma$.  The
solid line represents the lower limit as determined by the NC, and the
dotted line the upper limit as given by the WSC. For
$\gamma=(0,1,2,3)$ the black dots connected by the dashed line
represent the upper limits obtained using the SSC, while the true
limits derived from the DFs are shown as black squares. The
long--dashed line represents the limit on the anisotropy radius in
presence of a dominant central BH.
\label{fig1}}

\figcaption[]{The dimensionless DF $\fst/\fn$ for the one--component 
$\gamma=1$ model [solid line, $Q(0)=\psis(0)$ on the abscissae], and
in the case of a dominant $\gamma=0$ halo
[$\lim_{\mu\to\infty}\mu^{3/2}\fst/\fn$, $Q(0)=\psih(0)$ on the
abscissae], when $\beta=0.2$ (dotted line) and $\beta=3$ (dashed
line).  In the upper panel the globally isotropic case is shown, while
in the lower panel the anisotropy radius is fixed at
$\ra=0.26\rs$. The normalization constant is
$\fn=G^{-3/2}\ms^{-1/2}\rs^{-3/2}$.
\label{fig2}}

\figcaption[]{The dimensionless DF $\fha/\fn$ 
for the one--component $\gamma=0$ model [solid line, $Q(0)=\psih(0)$
on the abscissae, $\fn=G^{-3/2}\mh^{-1/2}\rh^{-3/2}$], and in the case
of a dominant $\gamma=1$ halo [$\lim_{\mu\to 0}\mu^{-1}\fha/\fn$,
$Q(0)=\psis(0)$ on the abscissae, $\fn=G^{-3/2}\ms^{-1/2}\rs^{-3/2}$],
when $\beta=0.2$ (dotted line) and $\beta=5$ (dashed line). In the
upper panel the globally isotropic case is shown, while in the lower
panel the anisotropy radius is fixed at $\ra=0.65\rh$.
\label{fig3}}

\figcaption[]{The central value of the dimensionless isotropic 
[solid line, equation (36)] and anisotropic [dotted line, equation (37)] 
parts of the DF for the halo--dominated $\gamma=0$ model, as a
function of $\beta=\rh/\rs$, for $\beta\geq 1$.
\label{fig4}}

\figcaption[]{The minimum value for the anisotropy radius of 
both components of (1,0) models in the case of a dominant halo, as a
function of $\beta=\rh/\rs$. The anisotropy radius is normalized to
the scale--length of the specific component.  In panel (a) the solid
line refers to the $\gamma=1$ component, and the dotted line to the
$\gamma=0$ component.  The two dashed lines represent the same
quantities, when the halo mass is ten times the mass of the specific
component.  The black dots are the critical values of $\sa$ for the
one--component models.  Panel (b) is the enlargement of the small
window at the top--right of panel (a). The dashed line shows the upper
limit on $\sa$ due to the halo concentration effect.
\label{fig5}}

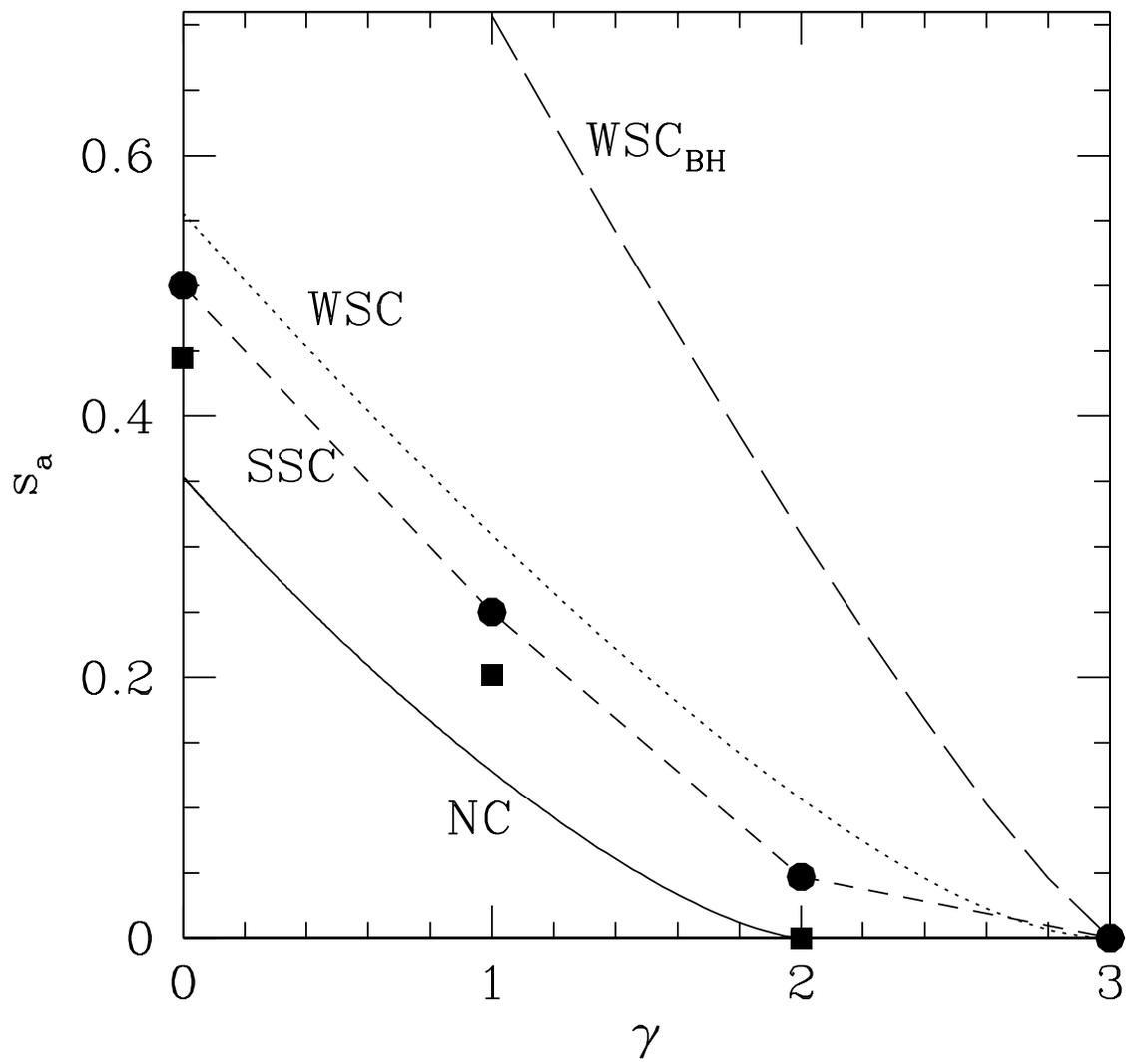
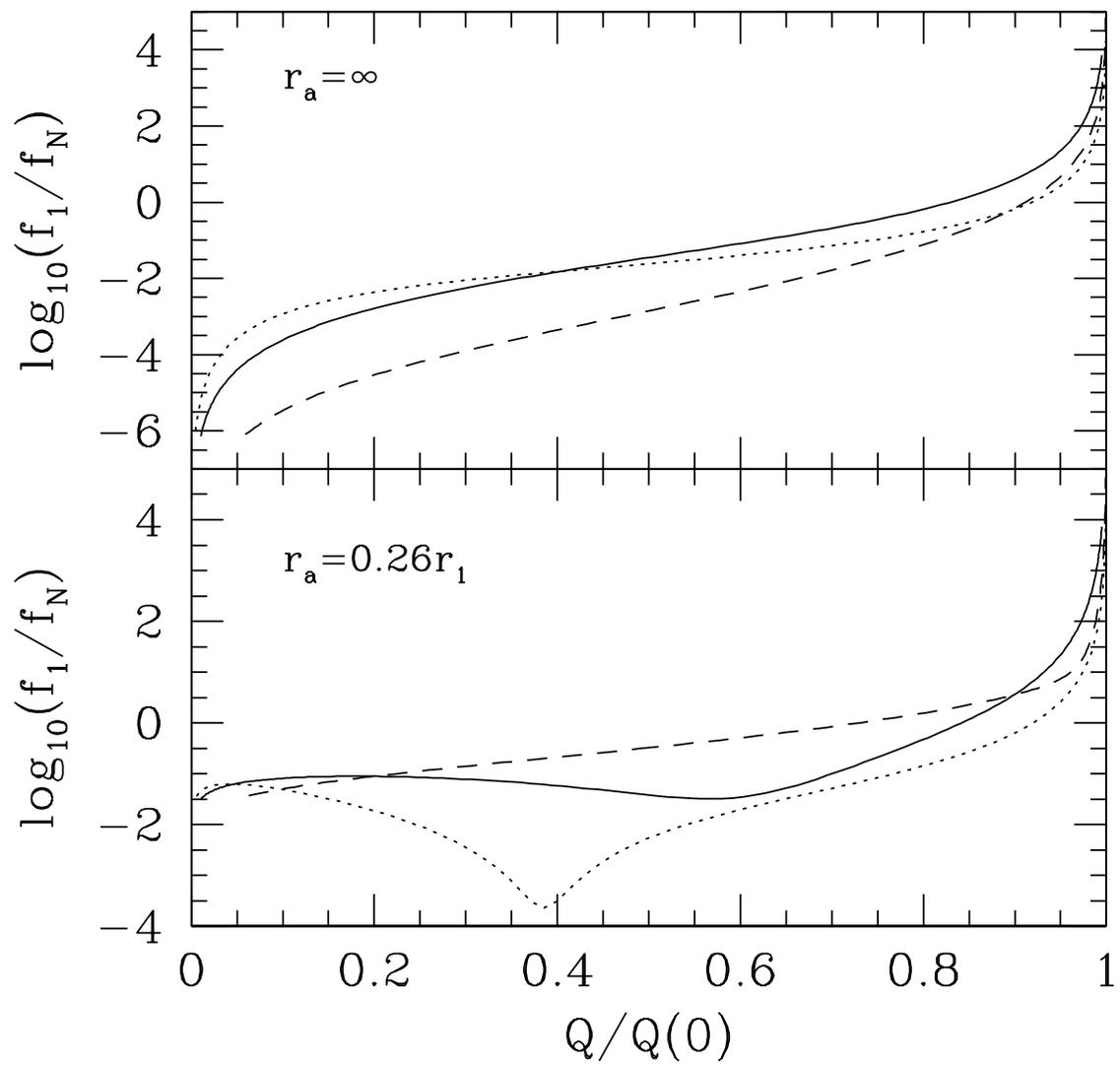
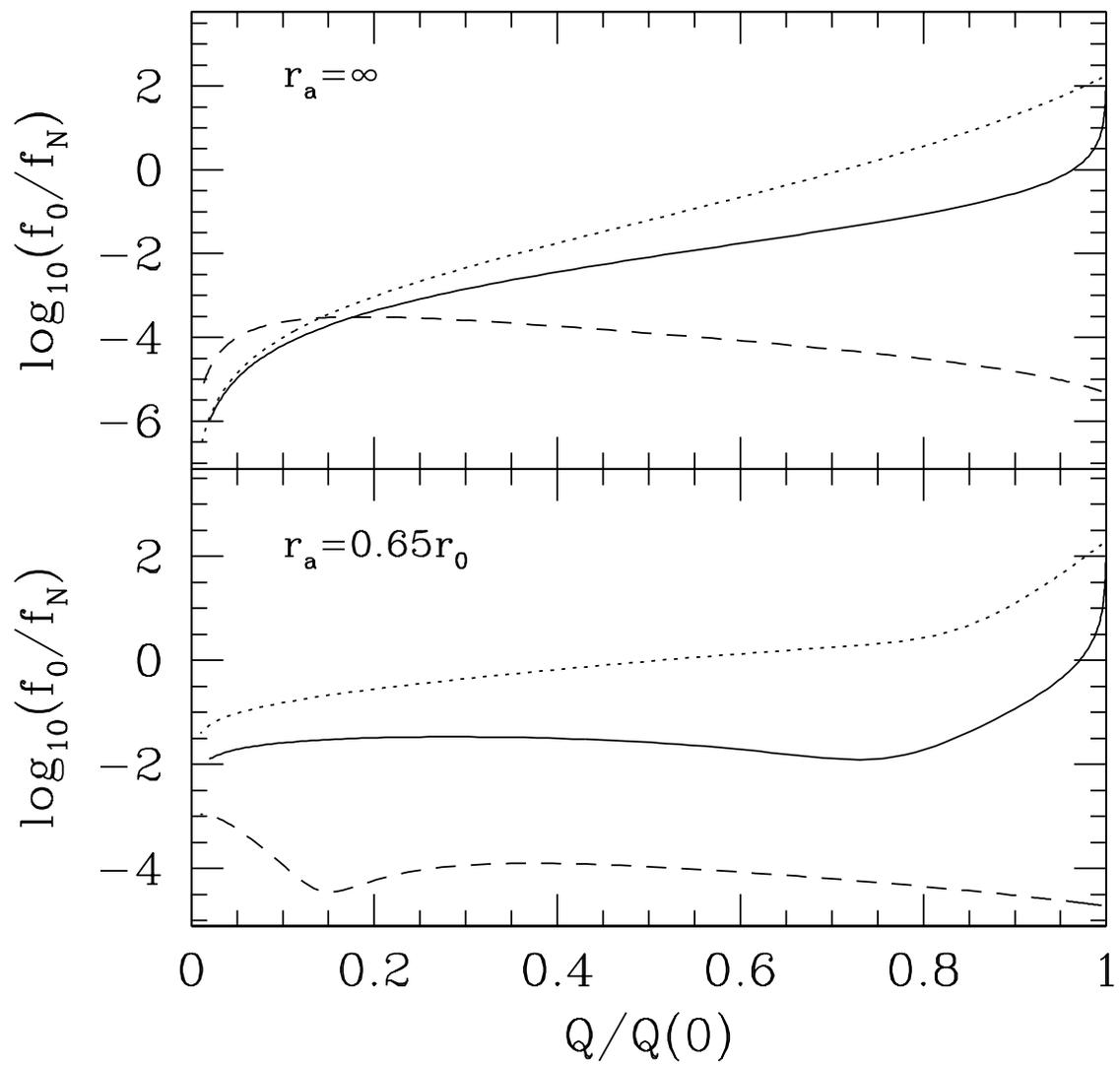
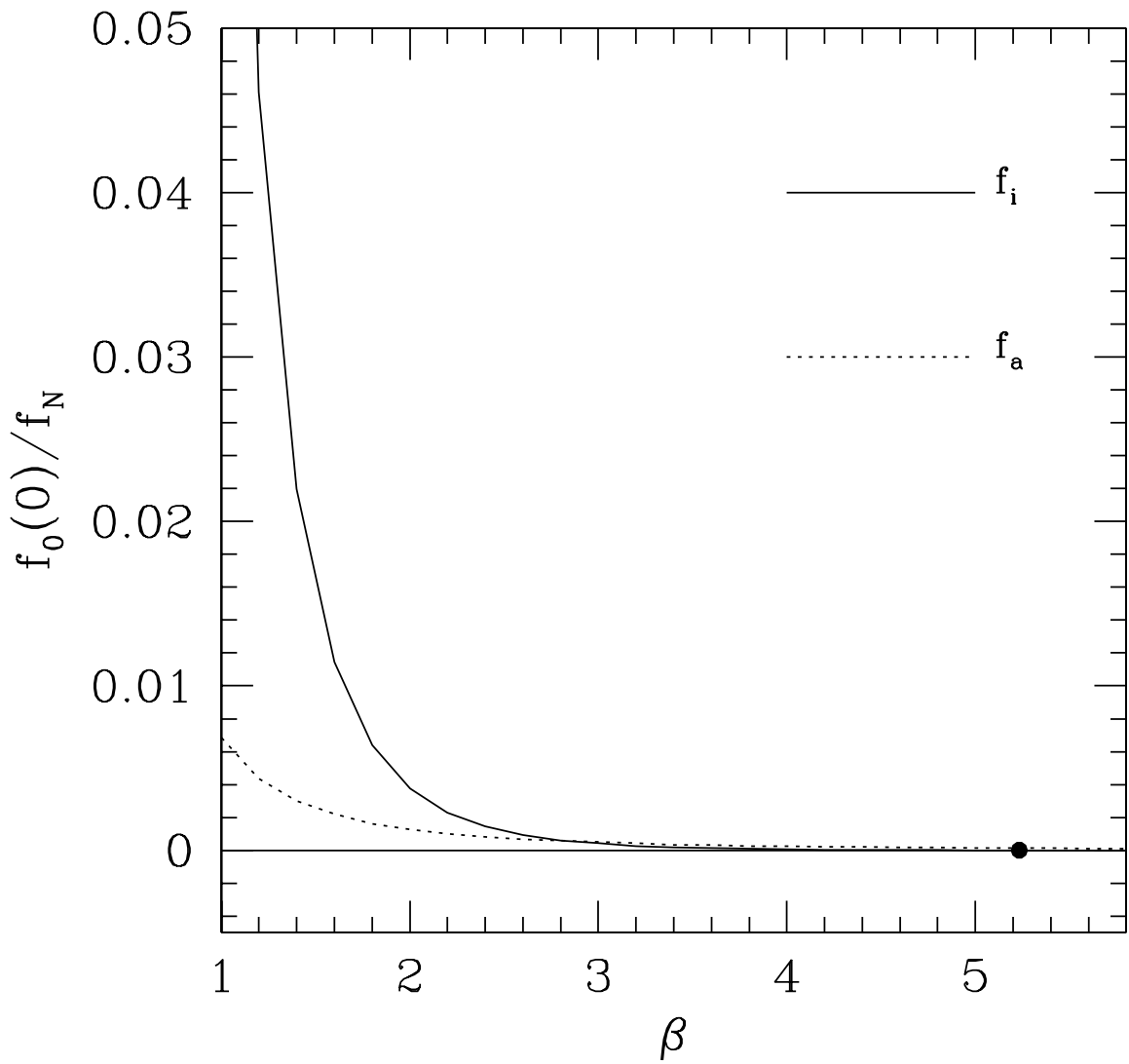
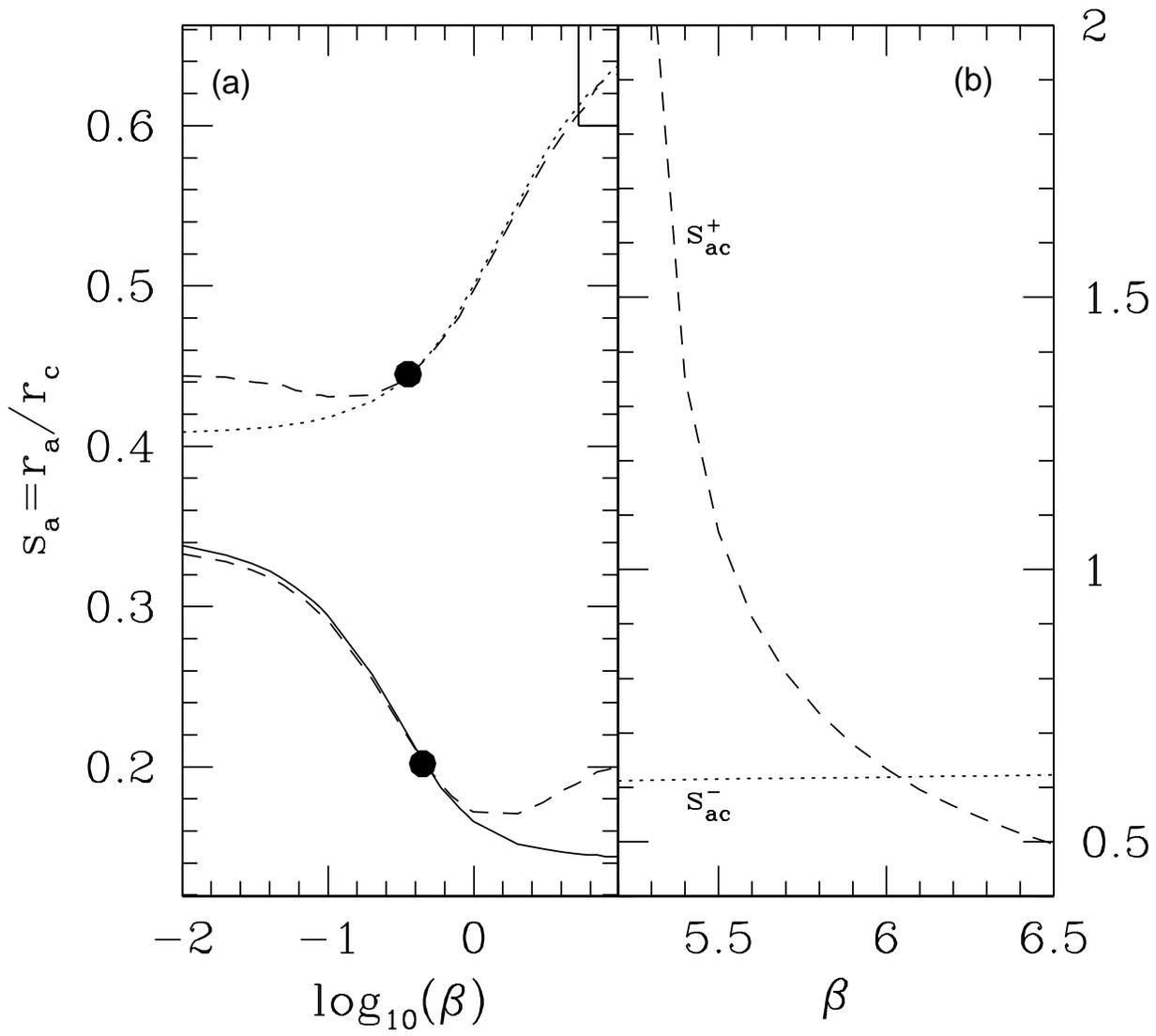
\end{document}